\documentclass[aps,preprint,superscriptaddress,showpacs,floatfix]{revtex4}%
\usepackage{amsfonts}
\usepackage{amsmath}
\usepackage{amssymb}
\usepackage{graphicx}
\usepackage{graphicx}%
\begin{document}
\title{Pulse shaping by a frequency filtering of a sawtooth phase-modulated CW laser}
\author{R. N. Shakhmuratov}
\affiliation{Zavoisky Physical-Technical Institute, FRC Kazan Scientific Center of
RAS, Kazan 420029 Russia}
\affiliation{Kazan Federal University, 18 Kremlyovskaya Street, Kazan 420008 Russia}
\date{{ \today}}

\begin{abstract}
The spectrum of a CW field whose phase experiences a periodic sawtooth
modulation is analyzed. Two types of the sawtooth phase modulation are
considered. One is created by combining many harmonics of the fundamental
frequency. The second is produced by electro-optic modulator fed by the
relaxation oscillator, which generates a voltage slowly rising during charging
the energy storage capacitor and dropping fast due to discharge by a short
circuit. It is proposed to filter out the main spectral component of the
sawtooth phase modulated field. This filtering produces short pulses from CW
phase modulated field. Several filters are proposed to remove the selected
spectral component. It is shown that such a filtering is capable to produce a
train of short pulses. Duty cycle of this train is equal to the modulation period
and duration of the pulses can vary from $10^{-1}$ to $10^{-2}$ of the period.
Depending on the modulation frequency, the proposed method is capable to
produce pulses with duration ranging from nanoseconds to a fraction of picosecond.

\end{abstract}
\maketitle

\section{Introduction}

Ultrafast optics is applied in widespread domains including but not limited to
high-field laser matter interactions, ultrafast time-resolved spectroscopy,
high precision frequency metrology and development of optical clocks,
nonlinear microscopy, optical communications, see for example Refs.
\cite{Book1,Book2}. Initially, short periodic pulses are generated by
high-repition-rate mode-locked lasers. However, in this method the generated
pulses and their corresponding spectral lines can suffer from instability
problems. Alternative passive systems generating pedestal-free optical pulses
with high peak power from a low-power laser employ large variety of methods to
compress the pulses. Among them one can mention, for example, acousto-optic
modulators (AOM) \cite{Warren1,Warren2,Verluise}, frequency chirping followed
by dispersive compensators
\cite{Treacy,Grischkowsky1974,Pearson,Nakatsuka,Nikolaus}, dispersive
modulators \cite{Loy1,Loy2}, high r.f. power spatial modulation of the field
phase by electro-optic modulators (EOM) followed by a lens
\cite{Kobayashi2005}, and modification of phases and amplitudes of the
spectral components of the phase modulated field by the programmable filters
to engineer the desired spectrum of the field \cite{Weiner2011}. Passive
systems using phase modulated CW laser offer several advantages. Among them
are lower cost and complexity, easy tuning of the frequency comb offset,
continuous tunability of the duty cycle, and reasonably stable operation
without active control. Phase modulating techniques, mentioned above, are
based on the phase manipulation of the spectral components of the frequency
comb, which makes phasing of these components. Therefore, duration of the
compressed pulses shortens with increase of the value of the phase modulation
index, while duty cycle is always equal to the modulation period.

A different method of pulse compression was recently reported in
\cite{Vagizov,Shakhmuratov2015,Shakhmuratov2017,Kocharovskaya2018}. The
capabilities of this method was experimentally demonstrated for gamma photons
with long duration of a single-photon wave packet
\cite{Vagizov,Shakhmuratov2015}. Splitting of a single-photon long pulse into
short pulses can be used to create time-bin qubits, whose concept was
introduced before in quantum informatics for optical photons
\cite{Gisin1999,Gisin2002}.

The method \cite{Vagizov,Shakhmuratov2015} is also based on frequency
modulation of the radiation field. However, in spite of subsequent phasing of
the produced spectral components, absorption (removal) of a particular
spectral component is used \cite{Vagizov,Shakhmuratov2015,Shakhmuratov2017}.
This removal method has many in common with other methods used before. The
method is also flexible and allows fine control of the duration and repetition
rate of the pulses. An appreciable shortening of the pulses
\cite{Shakhmuratov2015,Shakhmuratov2017} is achieved for high phase-modulation
index as in the previous passive methods of the pulse shaping.

In this paper, new modification of the phase modulation with subsequent
removal of the one of the spectral components of the comb is proposed. In this
variant, appreciable pulse compression can be achieved for moderate value of
the modulation index, which is even smaller than that produced by EOM fed by
the half-wave voltage $V_{\pi}$.

The core idea of the method is a sawtooth phase modulation in which the phase
periodically ramps upward and then sharply drops. Linear phase rise and sharp
drop are proposed to construct using additive synthesis of many harmonics of
frequency $\Omega$ with decreasing amplitudes according to the law $1/n$,
where $n$ is the number of the harmonic $n\Omega$. The larger the number $N$
of the highest harmonic, which is $N\Omega$, the sharper the phase drop is.
This simple model of the phase modulation allows to describe analytically all
the details of the pulse shaping based on one spectral component removal in
the spectrum of the phase modulated field. Duration of the compressed pulse,
predicted by the model, is $2(2N+1)$ times shorter than the period $T=2\pi
/\Omega$ of the phase modulation. Understanding of physics of this pulse
shaping technique allows to extend the method to the case of nonideal sawtooth
phase modulation with \ periodic nonlinear phase rise and exponential phase
drop. The faster the phase drops, the shorter the pulse is produced.

The sawtooth phase modulation technique is also based on the removal of one
spectral component of the comb spectrum of the phase modulated field similar
to the method of harmonic phase modulation
\cite{Vagizov,Shakhmuratov2015,Shakhmuratov2017}. This filtering is proposed
to implement by cloud of cold atoms, atomic vapors, organic molecules doped in
a polymer matrix, and liquid crystal phase modulator (LCM). Depending on the
value of the modulation frequency and selected frequency filter, one can
generate a sequence of pulses ranging from nanoseconds to a fraction of picosecond.

The paper is organized as follows. In Sec. II, a sawtooth phase modulation I,
created by mixing many harmonics of the fundamental frequency, is considered.
In Sec. III, frequency filtering of the phase modulated field is discussed.
In Sec. IV, periodic sawtooth phase modulation, which consists of slowly
rising stage according to the law $(1-e^{-t/T_{R}})$ and fast dropping stage
according to $e^{-t/T_{D}}$, and frequency filtering of the phase modulated
field are considered. Frequency filtering methods are discussed in Sec. V.

\section{Sawtooth phase modulation I}

We consider CW radiation field $E(t)=E_{0}\exp(-i\omega_{r}t+ikz)$, which
after passing through the electro-optic modulator acquires a sawtooth phase
modulation%
\begin{equation}
E_{EO}(t)=E(t)e^{i\varphi(t)}, \label{Eq1}%
\end{equation}
where
\begin{equation}
\varphi(t)=\sum_{n=-\infty}^{+\infty}\left(  \Omega t-2\pi n\right)  \left\{
\theta\left[  t-T\left(  n-\frac{1}{2}\right)  \right]  -\theta\left[
t-T\left(  n+\frac{1}{2}\right)  \right]  \right\}  , \label{Eq2}%
\end{equation}
$\Omega$ and $T=2\pi/\Omega$ are the modulation frequency and period, $n$ is
an integer varying from $-\infty$ to $+\infty$, and $\theta(x)$ is the
Heaviside step function. This kind of phase modulation is shown in Fig. 1 by
dotted blue line.

Physically, it is difficult to make instantaneous phase drop after a linear
ramp up. This problem can be solved by using additive synthesis of many
harmonics of frequency $\Omega$ with decreasing amplitudes. Fourier
transforms,
\begin{equation}
\frac{1}{T}\int_{-T/2}^{T/2}\varphi(t)e^{-in\Omega t}dt=i\frac{(-1)^{n}}{n},
\label{Eq3}%
\end{equation}
for $n\neq0$ and $\int_{-T/2}^{T/2}\varphi(t)dt=0$ for $n=0$, give the
frequency content of $\varphi(t)$ for the limited number $N$ of the harmonics,
i.e.,%
\begin{equation}
\varphi_{N}(t)=2\sum_{n=1}^{N}(-1)^{n+1}\frac{\sin(k\Omega t)}{n}, \label{Eq4}%
\end{equation}
where $N$ defines the highest frequency $N\Omega$ of the Fourier content of
the synthesized periodic phase evolution. Example of $\varphi_{N}(t)$ with
$N=5$ is shown in Fig. 1 by red solid line. It is interesting to notice that
modulation index of the main spectral component with $n=1$ is equal $2$, while
together with other four components the maximum phase shift is $\pi$, which
can be produced by the half-wave voltage $V_{\pi}$ applied to EOM.
\begin{figure}[ptb]
\resizebox{0.8\textwidth}{!}{\includegraphics{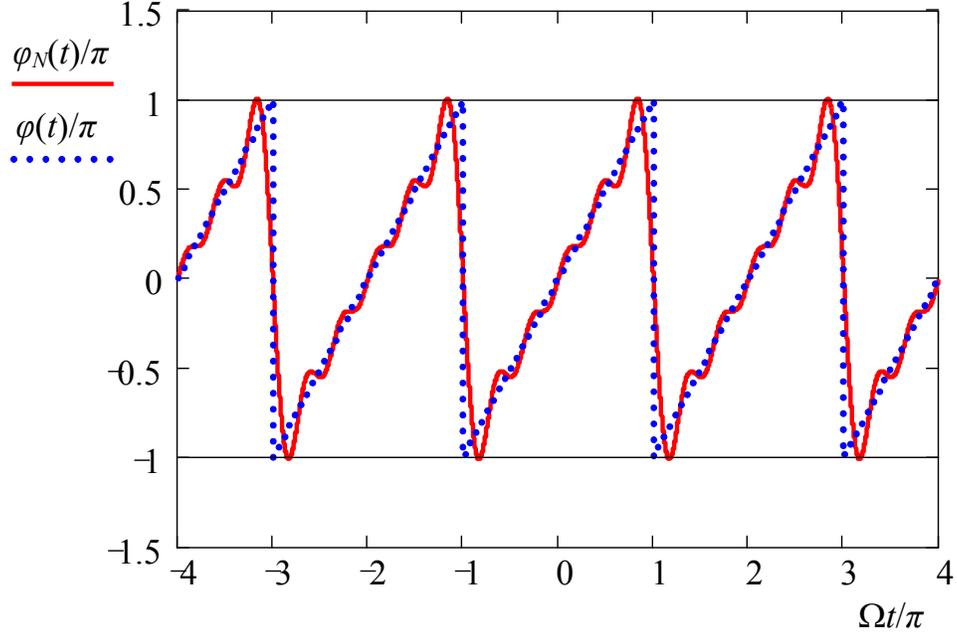}}\caption{Time
evolution of the sawtooth phase $\varphi(t)$ (blue dotted line) and the phase
$\varphi_{5}(t)$, synthesized from 5 harmonics (solid red line). Both are
normalized to $\pi$. Horizontal black bars show limiting values of the phase
change, $\pm\pi$}%
\label{fig:1}%
\end{figure}\begin{figure}[ptbptb]
\resizebox{0.8\textwidth}{!}{\includegraphics{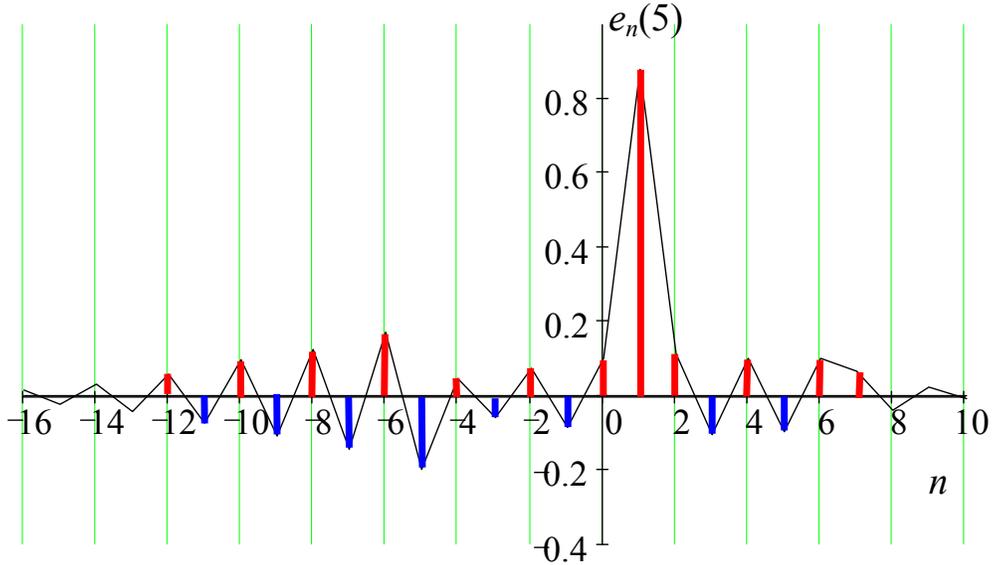}}\caption{Frequency
content of the field with sawtooth phase modulation synthesized from $N=5$
harmonics. The amplitudes of the spectral components are shown by red
(positive amplitudes) and blue (negative amplitudes) bars. Their maxima are
linked by black solid lines for visualization.}%
\label{fig:2}%
\end{figure}

Fourier transform%
\begin{equation}
\frac{1}{T}\int_{-T/2}^{T/2}e^{i\varphi_{N}(t)-in\Omega t}dt=e_{n}(N),
\label{Eq5}%
\end{equation}
allows to find the Fourier content of the field $E_{EO}(t)=E(t)e^{i\varphi
_{N}(t)}$ transmitted through AOM, which can be expressed as%
\begin{equation}
E_{EO}(t)=E_{0}e^{-i\omega_{r}t+ikz}\sum_{n=-\infty}^{n=\infty}e_{n}%
(N)e^{in\Omega t}. \label{Eq6}%
\end{equation}
The amplitudes of the spectral components $e_{n}(N)$ are shown in Fig. 2. The
component with $n=1$ has the largest amplitude. For example, for $N=5$ this
amplitude is $e_{1}(5)=0.879$. With increase of $N$, this amplitude tends to
$1$, i.e., $e_{1}(10)=0.936$ for $N=10$ and $e_{1}(50)=0.987$ for $N=50$.
Numerical analysis shows that $e_{1}(N)$ can be approximated as $e_{1}%
(N)\approx1-1.33/(2N+1)$.

The satellites of the component with $n=1$ have smaller amplitudes, and their
signs change such that nearest components to the central one, which we denote
as $n_{c}=1$, are positive ($n_{c}\pm1$), the next pair ($n_{c}\pm2$) is
negative, then the next amplitudes with numbers $n_{c}\pm3$ are positive, etc.
until $n_{c}\pm N$, see Fig. 2. The amplitudes of the components in each pair
are not equal, i.e., $e_{0}(5)=0.102$ and $e_{2}(5)=0.111$ for the nearest pair,
$e_{-1} (5)=-0.085$ and $e_{3}(5)=-0.105$ for the next pair. With increase
of $N$, the absolute values of the amplitudes of the satellites decrease, while
the number of satellites with noticeable value of the amplitudes increases
resulting in the spectrum broadening of the field. In addition, the amplitudes
of the pairs with numbers $n_{c}\pm1$ and $n_{c}\pm2$ tend to be equal for large
$N$, i.e., $e_{2}(50)=e_{0}(50)=0.013$ and $e_{3}(50)=e_{-1}(50)=-0.013$

Thus, sawtooth phase modulation makes a frequency offset of the CW field
$E(t)$ changing its carrier frequency to $\omega_{r}-\Omega$. This is the
first difference between the sawtooth phase modulation $\varphi_{N}(t)$, Eq.
(\ref{Eq4}), and harmonic phase modulation $\varphi_{h}(t)=m\sin\Omega t$. The
latter does not change the central frequency of the field. Here, $m$ is the
modulation index of the harmonic phase modulation. The second difference
between them is the qualitatively different dependence of the amplitudes of
the sidebands on their numbers. The amplitudes of the harmonics $+n\Omega$ and
$-n\Omega$, produced by the harmonic phase modulation, have the same sign if
$n$ is even, and they have opposite signs if $n$ is odd. The amplitudes of the
harmonics $(n_{c}+n)\Omega$ and $(n_{c}-n)\Omega$, produced by the sawtooth
phase modulation, have always the same sign for $n\leq N$. Here, we remind that
$n_{c}=1$ is the number of the central component of the frequency comb created
by the sawtooth phase modulation.

\section{Frequency filtering of the phase modulated field I}

If we selectively remove the central component $n_{c}$ of the phase modulated
field without change of all other spectral components, we expect that
remaining $2N$ components will phase and rephase with the period $T$. To explain
this point we just consider, for example, the interference of only two nearest
spectral pairs of the central components,%
\begin{equation}
E_{tp}(t)=E(t)\left[  e_{-1}(N)e^{-i\Omega t}+e_{0}(N)+e_{2}(N)e^{2i\Omega
t}+e_{3}(N)e^{3i\Omega t}\right]  , \label{Eq7}%
\end{equation}
and assume that $e_{0}(N)=e_{2}(N)=-e_{-1}(N)=-e_{3}(N)=a$, which is the case
when $N=50$ and $a=0.013$. Then, Eq. (\ref{Eq7}) can be expressed as%
\begin{equation}
E_{tp}(t)=2aE(t)e^{-i\Omega t}\left(  \cos\Omega t-\cos2\Omega t\right)  .
\label{Eq8}%
\end{equation}
This equation shows that when $\Omega t=2k\pi$ the amplitude $E_{tp}(t)$ is
zero because of destructive interference of the spectral pair $e_{0}(N)$,
$e_{2}(N)$ with the pair $e_{-1}(N)$, $e_{3}(N)$. When $\Omega t=(2k+1)\pi$
the amplitude $E_{tp}(t)$ is equal to $-4aE_{0}$ due to constructive
interference of the spectral pairs. Here, $k$ is an integer.

Analysis of the interference of all the pairs of the frequency comb, which is
described by Eq. (\ref{Eq6}) with removed central component $n_{c}$, is quite
complicated. Essential simplification is achieved if one employs the method
proposed in Refs. \cite{Shakhmuratov2015,Shakhmuratov2017}. It suggests to
express the filtered comb as%
\begin{equation}
E_{f}(t)=E(t)\left(  \sum_{n=-\infty}^{n=\infty}e_{n}(N)e^{in\Omega
t}-e_{n_{c}}(N)e^{in_{c}\Omega t}\right)  \label{Eq9}%
\end{equation}
or%
\begin{equation}
E_{f}(t)=E(t)\left[  e^{i\varphi_{N}(t)}-e_{n_{c}}(N)e^{in_{c}\Omega
t}\right]  .\label{Eq10}%
\end{equation}
Then, the intensity of the filtered field $I_{f}(t)=\left\vert E_{f}%
(t)\right\vert ^{2}$ can be presented as%
\begin{equation}
I_{f}(t)=I_{0}\left\{  1-2e_{n_{c}}(N)\cos\left[  n_{c}\Omega t-\varphi
_{N}(t)\right]  +e_{n_{c}}^{2}(N)\right\}  ,\label{Eq11}%
\end{equation}
where $I_{0}=E_{0}^{2}$. According to this expression the evolution of the
phase $\psi(t)=n_{c}\Omega t-\varphi_{N}(t)$ fully defines the interference of
the spectral pairs, mentioned above. Formally, this interference can be
considered as an interference of the whole comb, $E_{EO}(t)$, with the removed
component whose phase is changed by $\pi$, i.e., with the field $E_{sc}%
(t)=-e_{n_{c}}(N)E(t)e^{in_{c}\Omega t}$. In the case of atomic or molecular
filters, the field $E_{sc}(t)$ has a physical meaning. This field is
coherently scattered in forward direction by atoms whose resonant frequency is
$\omega_{r}-\Omega$
\cite{Shakhmuratov2015,Shakhmuratov2017,Shakhmuratov2011,Shakhmuratov2012}. In
the case of the programmable filters \cite{Weiner2011}, the field $E_{sc}(t)$
is virtual. \begin{figure}[ptb]
\resizebox{0.8\textwidth}{!}{\includegraphics{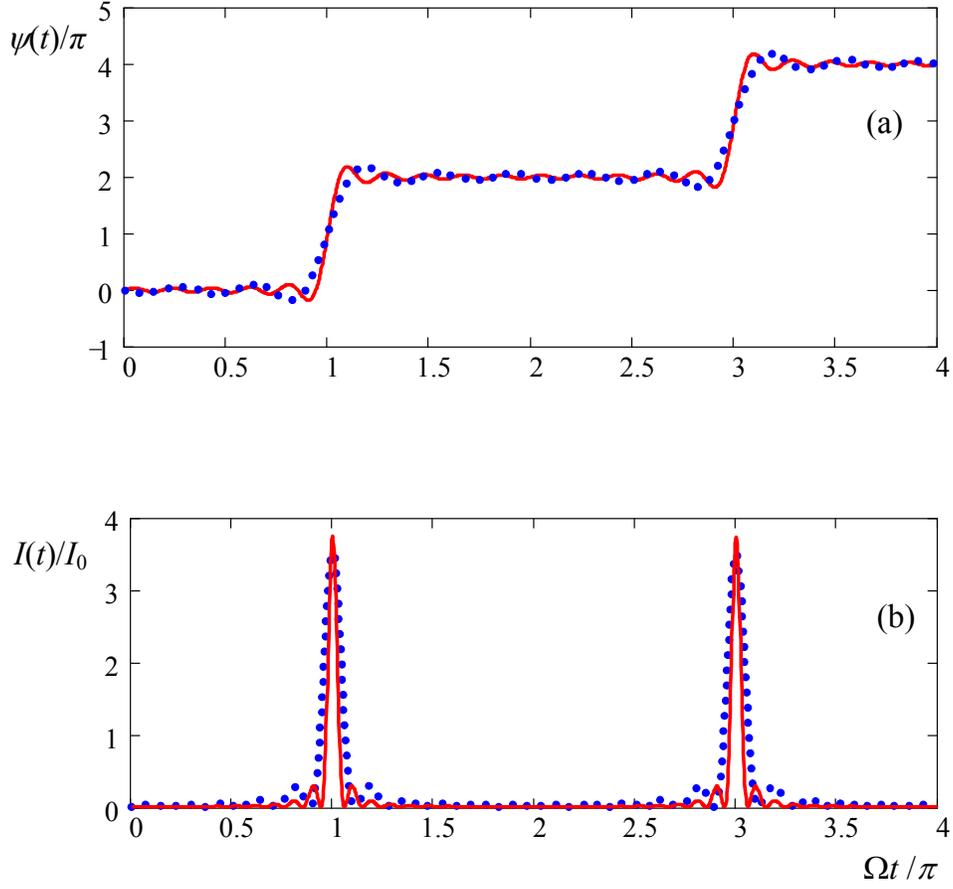}}\caption{(a) Time
evolution of the phase $\psi(t)=n_{c}\Omega t-\varphi_{N}(t)$, which specifies
the interference of the comb with the scattered field $E_{sc}(t)$. (b) Time
dependence of the intensity of the pulses normalized to $I_{0}$. Solid red
lines correspond to the sawtooth phase modulation with $N=10$, and blue dots
to $N=5$.}%
\label{fig:3}%
\end{figure}

The scattered field $E_{sc}(t)$ is in phase with the field $E_{EO}(t)$ when
$\psi(t)=(2k+1)\pi$, where $k$ is integer. Constructive interference of these
fields produces a pulse with intensity $I_{\max}=I_{0}[1+e_{n_{c}}(N)]^{2}$.
When $\psi(t)=2k\pi$, destructive interference of the fields results in the
drop of intensity to the level $I_{\min}=I_{0}[1-e_{n_{c}}(N)]^{2}$.
Substantial contrast between the pulse maximum and the pedestal is achieved if
$e_{n_{c}}(N)\longrightarrow1$. For example, for the sawtooth phase
modulation, synthesized from $5$ harmonics, we have $e_{1}(5)=0.879$, which
gives $I_{\min}=3.53I_{0}$ and $I_{\min}=0.015I_{0}$. Thus, for $N=5$, there
is a $23.7$-dB contrast ratio between the pedestal and the pulse maximum. For
$N=10$ and $N=50$, the contrast ratios are $30$ and $43.5$ dB, respectively.

Time evolution of the phase difference $\psi(t)$ of the comb $E_{EO}(t)$ and
coherently scattered field $E_{sc}(t)$ is shown in Fig. 3(a) for $N=5$ and
$10$. In time intervals $(k+1/2)T$ $<t<(k+3/2)T$, the phase difference
$\psi(t)$ is close to $2\pi(k+1)$, which results in destructive interference
of the fields. Here $k$ is an integer. On the borders of these time intervals
the phase difference $\psi(t)$ jumps from $2\pi(k+1)$ to $2\pi(k+2)$ crossing
the value $2\pi(k+3/2)$. At the crossing when $\psi(t)=2\pi(k+3/2)$, the pulse
is formed due to constructive interference of the fields, see Fig. 3(b). The
larger the number of harmonics $N$, the faster phase $\psi(t)$ crosses the
value $2\pi(k+3/2)$ and the shorter pulse is formed. The slope of the phase
change at the crossing point, which takes place at $t_{k}=(k+3/2)T$, is equal
to $\partial\psi(t)/\partial t\mid_{t_{k}}=(2N+1)\Omega$. Thus, the rate of
the crossing point is proportional to the modulation frequency $\Omega$ and
the number of harmonics $N$ constituting the sawtooth phase modulation.

The phase rise at the crossing point $t_{k}$ can be approximated as a linear
function $\psi(t)\approx\Omega t_{k}+(2N+1)\Omega(t-t_{k})$ with the slope
$\partial\psi(t)/\partial t\mid_{t_{k}}=(2N+1)\Omega$. This function fits well
the evolution of the phase $\psi(t)$ around the crossing points, see Fig. 4
where $t_{-1}$ crossing is shown. In the case $k=-1$, shown in the figure, the
linear approximation is reduced to $\psi(t)\approx(2N+1)\Omega t-2\pi N$. It
predicts that at $t_{-1}=T/2$ we have $\Omega t_{-1}=\pi$ and $\psi
(t_{-1})=\pi$. Then, constructive interference produces a field with maximum
intensity $I_{\max}=I_{0}[1+e_{n_{c}}(N)]^{2}$, which is $I_{\max}=4I_{0}$ if
$e_{n_{c}}(N)\longrightarrow1$. Neglecting the pedestal $I_{\min}%
=I_{0}[1-e_{n_{c}}(N)]^{2}$, one can roughly estimate from Fig. 4 that
intensity of the pulse drops to its half value $I_{\max}/2$ when $t=t_{-1}\pm
t_{\mathrm{half}}$ where $t_{\mathrm{half}}$ satisfies the relation
$(2N+1)\Omega t_{\mathrm{half}}=\pi/2$. At times $t=t_{-1}-t_{\mathrm{half}}$
and $t=t_{-1}+t_{\mathrm{half}}$ the phase $\psi(t)$ takes the values $\pi/2$
and $3\pi/2$, respectively. Intensity of the pulse drops two times at these
moments since $\cos\psi(t)$ in Eq. (\ref{Eq11}) is zero. Thus, the width of
the pulse can be estimated as $2t_{\mathrm{half}}=T/2(2N+1)$, i.e., it is
$(2N+1)$ time shorter than the half of the period $T=2\pi/\Omega$ of the phase
modulation. Taking into account that $e_{n_{c}}(N)$ is slightly smaller than
one, we numerically found that $t_{\mathrm{half}}$ satisfies slightly
different relation, which is $(2N+1)\Omega t_{\mathrm{half}}=\pi/1.923$. It
does not deviate significantly from our rough estimation.

\section{Sawtooth phase modulation II}

The sawtooth phase modulation can be also realized by EOM fed by a sawtooth
voltage, which is produced by relaxation oscillators. In this type of
oscillator the energy storage capacitor is charged slowly but discharged
rapidly by a short circuit through the switching device. Then, the ramp vltage
can be described by equation\begin{figure}[ptb]
\resizebox{0.8\textwidth}{!}{\includegraphics{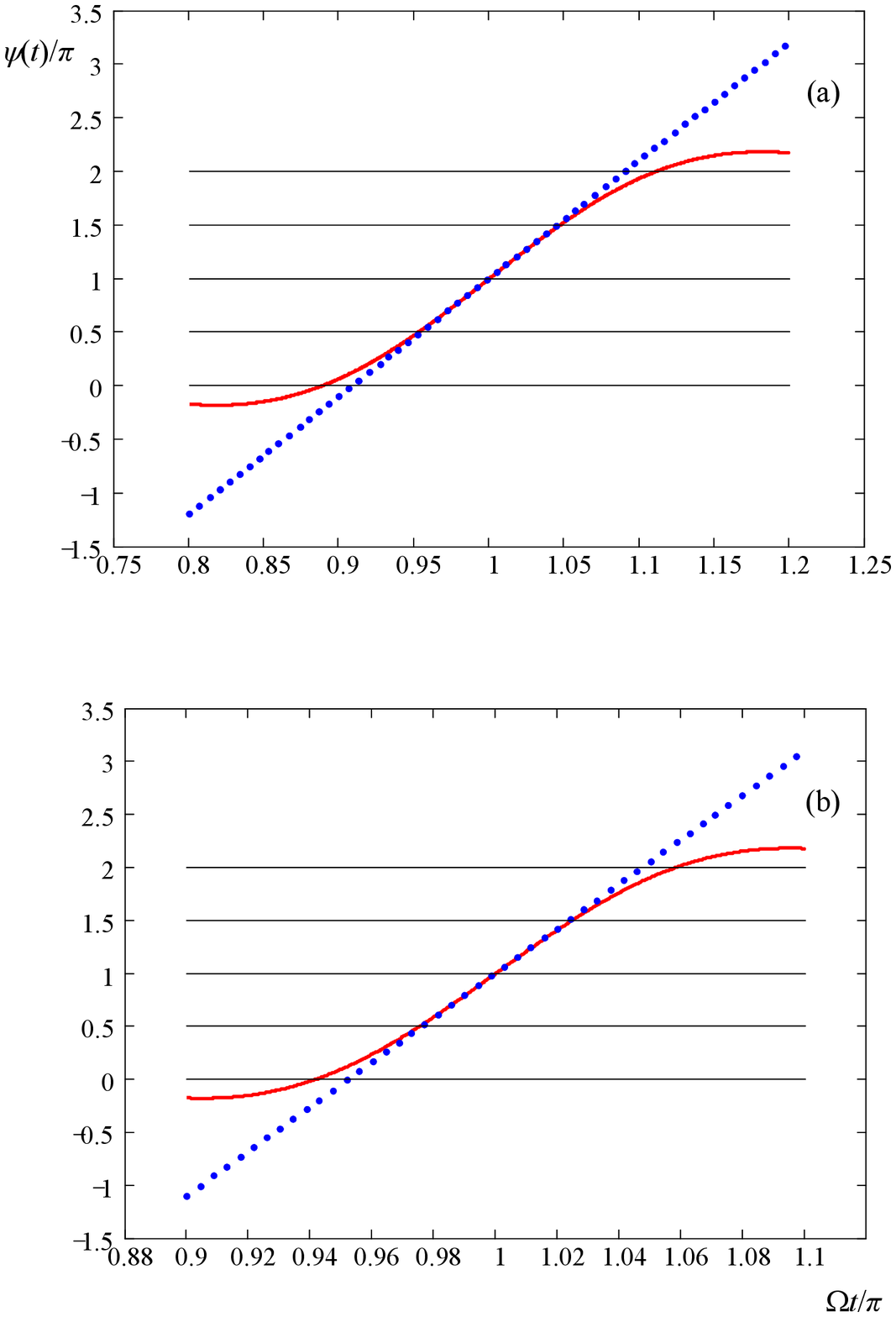}}\caption{Time
evolution of the phase $\psi(t)$ near one of the crossing points, i.e., at
$t_{k}$ with $k=-1$, is shown by red solid line. Its approximation by linear
time dependence is shown by blue dots (see the tex for details). The number of
harmonics constituting the sawtooth phase modulation is $N=5$ in (a) and 10 in
(b).}%
\label{fig:4}%
\end{figure}%
\begin{equation}
U_{R}(t)=U_{0}\left(  1-e^{-t/T_{R}}\right)  +U_{\min}, \label{Eq12}%
\end{equation}
where $U_{0}$ is a maximum voltage, $T_{R}$ is a rise time, and $U_{\min}$ is
an initial voltage, from which the ramp starts. The voltage drop is described
by%
\begin{equation}
U_{D}(t)=U_{\max}e^{-t/T_{D}}, \label{Eq13}%
\end{equation}
where $U_{\max}$ is a voltage when the discharge starts and $T_{D}$ is a drop time.

A periodic phase modulation, produced by such a sawtooth voltage, can be expressed
as follows%
\begin{equation}
\varphi_{RO}(t)=C\sum_{k=0}^{+\infty}\phi\left[  t-k\left(  T_{R}%
+T_{D}\right)  \right]  , \label{Eq14}%
\end{equation}
where%
\begin{equation}
\phi(t)=\phi_{R}(t)+\phi_{D}(t), \label{Eq15}%
\end{equation}%
\begin{equation}
\phi_{R}(t)=\left(  1-e^{-t/T_{R}}\right)  \left[  \theta(t)-\theta
(t-T_{R})\right]  , \label{Eq16}%
\end{equation}%
\begin{equation}
\phi_{D}(t)=\left[  e^{-(t-T_{R})/T_{D}}-e^{-1}\right]  \left[  \theta
(t-T_{R})-\theta(t-T_{R}-T_{D})\right]  . \label{Eq17}%
\end{equation}
Here, for simplicity, it is assumed that the rise and drop time periods are
equal to $T_{R}$ and $T_{D}$, respectively, and time independent part of the
phase is disregarded resulting in the condition $\phi(0)=0$. The maximum value
of the phase $\phi(t)$ at $t=T_{R}$ is taken equal to $2\pi$, which gives
$C=2\pi/\left(  1-e^{-t/T_{R}}\right)  $. The period of this sawtooth-phase
modulation is $T=T_{R}+T_{D}$. Time evolution of the phase $\varphi_{RO}(t)$
is shown in Fig. 5.

Fourier transform%
\begin{equation}
\frac{1}{T}\int_{0}^{T}e^{i\varphi_{RO}(t)-in\Omega t}dt=c_{n}, \label{Eq18}%
\end{equation}
allows to find the Fourier content of the field $E_{RO}(t)=E(t)e^{i\varphi
_{RO}(t)}$, which is%
\begin{equation}
E_{RO}(t)=E_{0}e^{-i\omega_{r}t+ikz}\sum_{n=-\infty}^{n=\infty}c_{n}%
e^{in\Omega t}. \label{Eq19}%
\end{equation}
\begin{figure}[ptb]
\resizebox{0.8\textwidth}{!}{\includegraphics{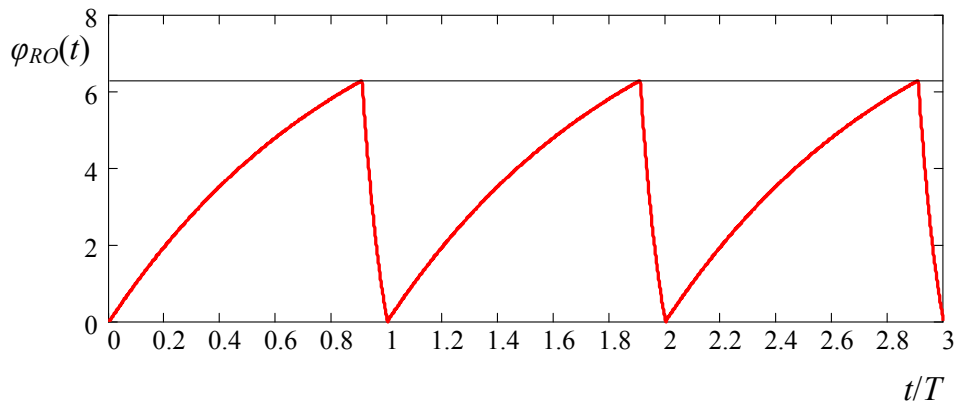}}\caption{Time
evolution of the phase $\varphi_{RO}(t)$. Time scale is normalized to
$T=T_{R}+T_{D}$. The parameters of the sawtooth phase modulation are related
as $T_{D}=T_{R}/10$. Black horizontal bar corresponds to $2\pi$.}%
\label{fig:5}%
\end{figure}where $\Omega=2\pi/T$.

Following the derivation method presented in Sec. III, one can obtain that
removing of the frequency component $\omega_{r}-\Omega$, whose amplitude is
$c_{1}E_{0}$, modifies the phase modulated field as%
\begin{equation}
E_{f}(t)=E(t)\left(  e^{i\varphi_{RO}(t)}-c_{1}e^{i\Omega t}\right)  ,
\label{Eq20}%
\end{equation}
whose intensity is
\begin{equation}
I_{f}(t)=I_{0}\left[  1-2a_{1}\cos\psi_{RO}(t)+a_{1}^{2}\right]  ,
\label{Eq21}%
\end{equation}
where
\begin{equation}
\psi_{RO}(t)=\Omega t+\xi_{1}-\varphi_{RO}(t). \label{Eq22}%
\end{equation}
Here $a_{1}$ and $\xi_{1}$ are the modulus and argument (phase) of the complex
number $c_{1}$, i.e. $c_{1}=a_{1}\exp(i\xi_{1})$.

Example of the formation of pulses is shown in Fig. 6(a) by blue dotted line
for the case when the phase drop is ten times faster than the phase rise,
i.e., for $T_{D}=T_{R}/10$. In this case $a_{1}=0.892$ and $\xi_{1}%
=0.808\approx\pi/3.9$. Absolute value $a_{1}$ of $c_{1}$ is close to unity.
Therefore the peak pulse intensity is $3.58$ (almost four) times larger than
the intensity of the CW field $I_{0}$. Evolution of the phase $\psi_{RO}(t)$,
which governs the interference of the incident field, $E_{EO}(t)$, with
scattered field, $E_{sc}(t)=-c_{1}E(t)e^{i\Omega t}$, is shown in Fig. 6(a) by
red solid line. Each time when $\psi_{RO}(t)$ crosses the value $(2k+1)\pi$,
the pulse is formed.

Zoom in on the area of the pulse formation around $t=T$ is shown in Fig. 6(b).
Numerical analysis gives an estimation of the pulse duration $t_{p}$ (full
width at half maximum), which is $0.041T_{R}$ for $T_{D}=T_{R}/10$. Thus,
during a short time of the phase drop $T_{D}$, the pulse is mainly formed
within the time interval $0.41T_{D}$, which is slightly less than a half of
$T_{D}$.
\begin{figure}[ptb]
\resizebox{0.8\textwidth}{!}{\includegraphics{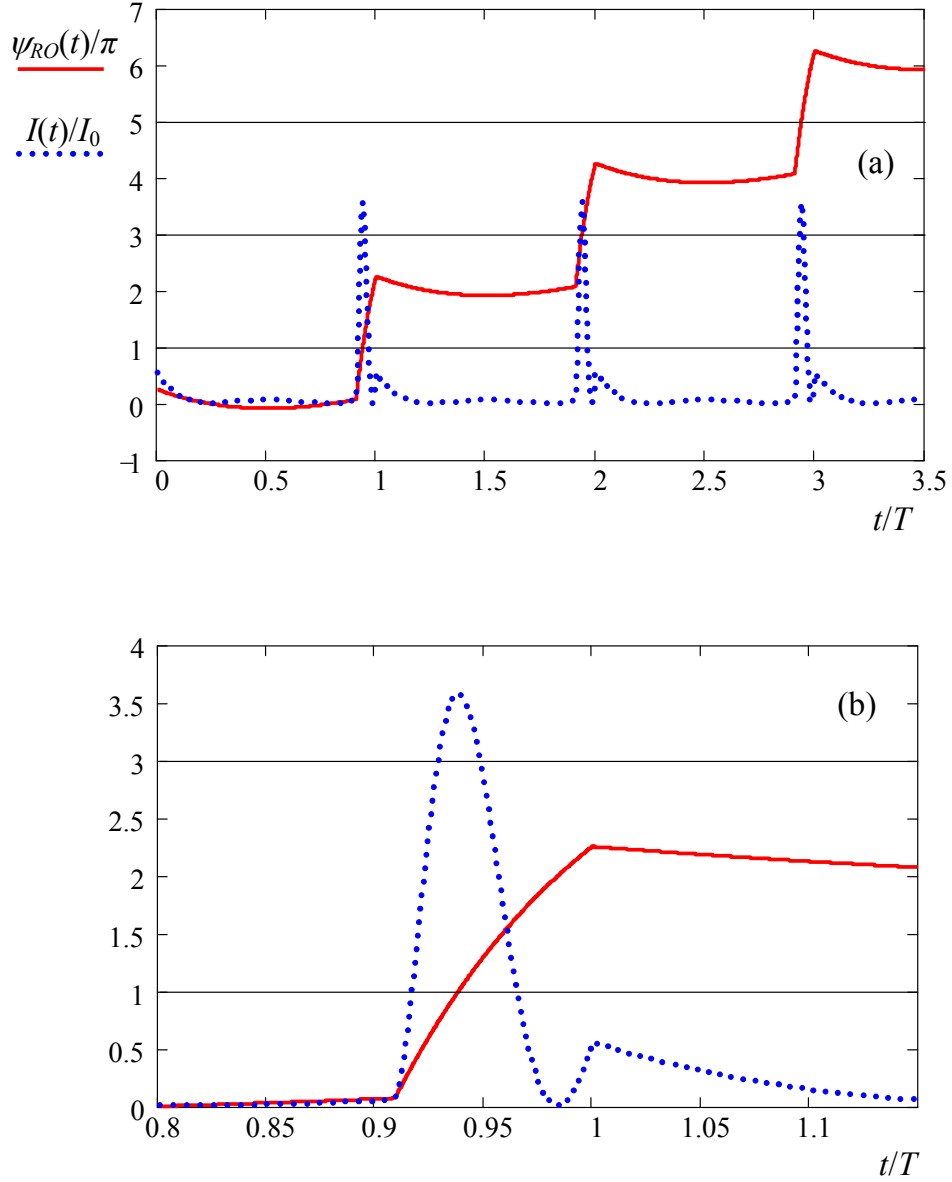}}\caption{(a) Time
evolution of the phase $\varphi_{RO}(t)/\pi$ (in units of $\pi$) is shown by
red solid line. Sequence of pulses, generated by the frequency filtering, is
shown by blue dotted line. The field intensity is normalized to $I_{0}$. Time
scale is in units of the period $T$. Both plots correspond to the case when
$T_{D}=T_{R}/10$. (b) Zoom in on the area of the pulse formation around $t=T$.
In both plots, horizontal black thin lines indicate the levels corresponding
to the phase values equal to $(2k+1)\pi$.}%
\label{fig:6}%
\end{figure}

\section{Frequency filtering methods}

Removal of the selected spectral component of the frequency comb, created by
the sawtooth phase modulation I and II, can be implemented by resonant filters
with a single absorption line $L(\omega-\omega_{f})$ centered at frequency
$\omega_{f}$. The width of this line, $\Gamma$, is to be much smaller than the
distance $\Omega$ between the frequency components of the comb, i.e.
$\Gamma\ll\Omega$.

If the line is homogeneously broadened, then optically thick absorber filters
out the selected frequency component $E_{s}(t)=E_{s}e^{-i\omega_{s}t+ikr}$
diminishing its amplitude as follows, see Refs. \cite{Shakhmuratov2017,Crisp},%
\begin{equation}
E_{fs}=\exp\left[  -\frac{\alpha_{B}l}{2}L(\omega_{s}-\omega_{f})\right]
E_{s},\label{Eq23}%
\end{equation}
where%
\begin{equation}
L(\omega_{s}-\omega_{f})=\frac{\Gamma/2}{\Gamma/2-i(\omega_{s}-\omega_{f}%
)}\label{Eq24}%
\end{equation}
is the Lorentzian profile describing the absorption and dispersion in the
filter of physical thickness $l$ and Beer's law absorption coefficient
$\alpha_{B}$. Here, $\Gamma$ is a width at half-maximum of the Lorentzian
absorption line of a single particle in the absorber. In exact resonance
($\omega_{s}=\omega_{f}$), the amplitude of the selected line decreases as
$E_{fs}=\exp\left(  -d/2\right)  E_{s}$, where $d=$ $\alpha_{B}l$ is the
optical thickness of the filter. The filtering becomes effective if $d/2\gg1$.
However, there is a limit set to the optical thickness of the filter by the
condition that the spectral components $\omega_{s}\pm\Omega$ neighboring the
selected component $\omega_{s}$ must be unaffected. This condition is
satisfied, if $d\ll4\Omega/\Gamma$, see Ref. \cite{Shakhmuratov2017}. The
amplitudes of the neighboring components, $E_{s\pm1}$, are changed according
to the equation%
\begin{equation}
E_{fs\pm1}=\exp\left[  -\frac{d}{2}L(\pm\Omega)\right]  E_{s\pm1}.\label{Eq25}%
\end{equation}
If $\Omega\gg\Gamma$, this change is mainly caused by the phase factor
$\exp(\pm id\Gamma/4\Omega)$ since $-\frac{d}{2}L(\pm\Omega)$ in the exponent
is approximated as $\pm id\Gamma/4\Omega$. The condition $d\ll4\Omega/\Gamma$
allows avoiding the modification of the neighboring components of the field spectrum.

If $\omega_{s}=\omega_{f}$ and only the selected frequency component is
affected by the resonant filter, then Eq. (\ref{Eq10}), describing in Sec. III
the filtered comb in the ideal case of $100\%$ removal/filtering, is modified
as
\begin{equation}
E_{f}(t)=E(t)\left[  e^{i\varphi_{N}(t)}-e_{n_{c}}(N)\left(  1-e^{-d/2}%
\right)  e^{in_{c}\Omega t}\right]  . \label{Eq26}%
\end{equation}
In a similar way, Eq. (\ref{Eq20}) in Sec. IV is modified in the case of the
resonant filter.

If absorption line in the filter is Doppler broadened, then the function
$L(\omega-\omega_{f})$ in Eq. (\ref{Eq25}) is replaced by%
\begin{equation}
F_{D}(\omega-\omega_{f})=\sqrt{\frac{\ln2}{\pi}}\frac{1}{\Delta\omega_{D}}%
\int_{-\infty}^{+\infty}L(\omega-\omega_{f}+\omega)e^{-\ln2\left(
2\omega/\Delta\omega_{D}\right)  ^{2}}\omega,\label{Eq27}%
\end{equation}
where $\Delta\omega_{D}$ is the Doppler width, which is supposed to be much
larger than $\Gamma$. At exact resonance we have $F_{D}(0)=\sqrt{\pi\ln
2}\Gamma/\Delta\omega_{D}$. Then, the amplitude of the component filtered by
the absorber with inhomogeneously broadened line decreases as $E_{fs}%
=\exp[-dF_{D}(0)/2]E_{s}$. The neighboring components are not affected if
$\Omega\gg\Delta\omega_{D}$. However, because for $\Gamma/\Delta\omega
_{D}=10^{-2}$ and $\Omega>1.8\Delta\omega_{D}$ the function $F_{D}%
(\omega-\omega_{f})$ has Lorentzian wings \cite{Shakhmuratov2008}, i.e., it
becomes $F_{D}(\omega-\omega_{f})=(\Gamma/2)/[(\Gamma/2)-i(\omega-\omega
_{f})]$, their influence on the spectral neighbors of the filtered component
is negligible if $d\ll4\Omega/\Gamma$. Thus, effective filtering takes place
if $\Omega\gg\Delta\omega_{D}$ and optical thickness satisfies the condition
$1.35\Delta\omega_{D}/\Gamma\ll d\ll$ $4\Omega/\Gamma$, see Ref.
\cite{Shakhmuratov2017}.

The filtering of the selected spectral component can be also implemented by
the method based on spectral line-by-line pulse shaper, see, for example, Ref.
\cite{Weiner1}. Many-pixel liquid crystal modulator (LCM) array allows in this
technique to control both amplitude and phase of individual spectral lines of
the field with a comb spectrum. LCM can be tuned such that only selected
spectral line of the comb is suppressed.

Effective and flexible method of creating nanosecond pulses can be implemented
by filtering of the frequency comb through laser-cooled atoms with a modest
optical depth. For example, $D_{1}$-line transition ($\lambda=795$ nm) of
$^{85}$Rb atoms in a two-dimensional magneto-optical trap have almost
homogeneous width $\Gamma\approx6$ MHz, see, for example, Ref. \cite{Chen}.
Therefore, with the modulation frequency of EOM $\Omega=30$ MHZ one can
generate $1.5$ ns pulses for the sawtooth phase modulation I with $N=5$ and $1.24$
ns\ for the sawtooth phase modulation II with $T_{D}=T_{R}/10$ by the filtering
through the cloud of laser-cooled $^{85}$Rb atoms. For $N=10$, pulse duration
shortens to $800$ ps. If the modulation frequency is increased to $\Omega=300$
MHZ, then the duration of the generated pulses is shortened to $150$ ps for
$N=5$ and to $80$ ps for $N=10$. For the sawtooth phase modulation II with
$T_{D}=T_{R}/10$ pulses shorten to $124$ ns.

As a frequency filter one can use a vapor of $^{87}$Rb atoms. Assume that the
selected frequency of the comb is tuned in resonance with the $S_{1/2}%
,F=1\rightarrow P_{1/2},F=2$ transition of the $D_{1}$ line of natural Rb
($\lambda=795$ nm). Below, we take the parameters of the experiment
\cite{Lukin} where spectral properties of the electromagnetically induced
transparency were studied in this vapor. Natural linewidth of the Rb $D_{1}$
line is $\Gamma=5.4$ MHz and Doppler broadening is $\Delta\omega_{D}=500$ MHz.
Selecting the phase modulation frequency $\Omega=10$ GHz, which is 20 times
larger than the Doppler width $\Delta\omega_{D}=500$ MHz, we satisfy the
condition $\Omega\gg\Delta\omega_{D}$. According to the estimates given in
Ref. \cite{Shakhmuratov2017} for the Rb cell with the length $l=5$ cm and
atomic density $N_{1}=6\times10^{10}$ cm$^{-3}$, the modification of the
spectral components neighboring the selected one is almost negligible. For
this atomic density the effective optical depth of the cell at the selected
line center is $dF_{D}(0)=14.4$ while $d=905$. With these values of the
parameters $\Omega$, $\Delta\omega_{D}$, $\Gamma$, and $d$ the condition
$1.35\Delta\omega_{D}/\Gamma\ll d\ll$ $4\Omega/\Gamma$ is easily satisfied.

For the modulation frequency $\Omega=10$ GHz, filtering through the atomic
vapor or removing of the selected spectral component with the help of LCM
\cite{Weiner1} produce much shorter pulses. For example, for the sawtooth
phase modulation I with $N=5$ and sawtooth phase modulation II with
$T_{D}=T_{R}/10$ one can generate $4.5$ and $3.7$ ps pulses, respectively. For
the sawtooth phase modulation I consisting of 10 harmonics ($N=10$), duration
of the pulses shortens to $2.4$ ps. If the number of the harmonics increases
to $N=50$, pulse duration shortens to $495$ fs.

As a selective filter one can use organic molecules doped in polymer matrix.
It is experimentally possible to burn a broad spectral hole in their spectrum
with a sharp absorption peak sitting at its center. Such a structure is
persistent at liquid helium temperature. The frequency resolution of the
persistent spectral hole burning is limited by the width of the homogeneous
zero-phonon line of the chromophore molecules, which typically has a width of
$10^{-2}-10^{-4}$ cm$^{-1}$ or less \cite{Rebane1995,Rebane2002}. The holes
could be burned in a planar waveguide geometry where a thin polymer film with
doped molecules is superimposed\ as a cover layer on a planar glass waveguide
\cite{Tschanz1995,Tschanz1996}. Then, illumination in the transverse direction
with low absorption creates a hole, while weak probing field propagates in a
longitudinal wave guiding direction with high absorption. For example, such a
waveguide with a spectral hole acting as subgigahertz narrow-band filter was
proposed to observe slow light phenomenon in Refs.
\cite{Shakhmuratov2005,Rebane2007}.

\section{Conclusion}

Pulse shaping by the spectral line pulse shaper is capable to produce short
pulses from the CW phase modulated field. Harmonic phase modulation with a
single frequency produces pulses or bunches of pulses with the duty factor
equal to the modulation period. Duration of the pulses shortens with increase
of the phase modulation index. Liquid crystal phase modulator (LCM) can
produce pulses whose duration is an order of magnitude shorter than the
modulation period if $\sim 10$ spectral components with noticeable amplitudes are
phased by LCM. Removal of the selected spectral component is also capable to
produce pulses of comparable duration. However, as the LCM method, it produces
such a short pulses if modulation index is larger than $\pi$. Moreover, with
increase of the modulation index the contrast between the pedestal and pulse
maximum decreases in the removal method. Pulse shaping by the removal of the
selected spectral component of the CW sawtooth phase modulated field works
with the fixed modulation index of moderate value. Short pulses are generated
during fast dropping of the phase. The faster this drop is, the shorter the
pulse is formed. Its duration can made an order or two orders of magnitude
shorter than the phase modulation period. The contrast between the pedestal
and the pulse maximum increases with increase of the rate of the phase drop.

\section{Acknowledgments}

This work was funded from the government assignment for FRC Kazan Scientific
Center of RAS

\end{document}